# Solenoidal and potential velocity fields in weakly turbulent premixed flames

Vladimir A. Sabelnikov[1,2], Andrei N. Lipatnikov[3], Nikolay Nikitin[4], Shinnosuke Nishiki[5], Tatsuya Hasegawa[6]

[1]*ONERA – The French Aerospace Laboratory, F-91761 Palaiseau, France*
[2]*Central Aerohydrodynamic Institute (TsAGI), 140180 Zhukovsky, Moscow Region, Russian Federation*
[3]*Department of Mechanics and Maritime Sciences, Chalmers University of Technology, Gothenburg, 41296 Sweden*
[4]*Moscow State University, Moscow Region, Russian Federation*
[5]*Department of Information and Electronic Engineering, Teikyo University, Utsunomiya 320-8551, Japan,*
[6]*Institute of Materials and Systems for Sustainability, Nagoya University, Nagoya 464-8603, Japan*

**Corresponding author:** Prof. Andrei Lipatnikov, Department of Mechanics and Maritime Sciences, Chalmers University of Technology, Gothenburg, 41296 Sweden, lipatn@chalmers.se

**Abstract**

Direct Numerical Simulation data obtained earlier from two statistically 1D, planar, fully-developed, weakly turbulent, single-step-chemistry, premixed flames characterized by two significantly different (7.53 and 2.50) density ratios $\sigma$ are analyzed to explore the influence of combustion-induced thermal expansion on the turbulence and the backward influence of such flow perturbations on the reaction-zone surface. For this purpose, the simulated velocity fields are decomposed into solenoidal and potential velocity subfields. The approach is justified by the fact that results obtained adopting (i) a widely used orthogonal Helmholtz-Hodge decomposition and (ii) a recently introduced natural decomposition are close in the largest part of the computational domain (including the entire mean flame brushes) except for narrow zones near the inlet and outlet boundaries. The results show that combustion-induced thermal expansion can significantly change turbulent flow of unburned mixture upstream of a premixed flame by generating potential velocity fluctuations. Within the flame brush, the potential and solenoidal velocity fields are negatively (positively) correlated in unburned reactants (burned products, respectively) provided that $\sigma = 7.53$. Moreover, correlation between strain rates generated by the solenoidal and potential velocity fields and conditioned to the reaction zone is positive (negative) in the leading (trailing, respectively) halves of the mean flame brushes. Furthermore, the potential strain rate correlates negatively with the curvature of the reaction zone, whereas the solenoidal strain rate and the curvature are negatively (positively) correlated in the leading (trailing, respectively) halves of the mean flame brushes. Finally, if $\sigma = 7.53$, the stretch rate conditioned to the reaction zone is negative (positive) in regions characterized by large positive (negative) potential strain rates, whereas the opposite trend is observed for the solenoidal strain rate. Thus, the potential and solenoidal velocity fields differently affect the reaction-zone-surface area. In the case of $\sigma = 2.5$, such differences are significantly less pronounced.

## 1. Introduction

Since a hypothesis of flame-generated turbulence was put forward by Karlovitz et al. [1] and Scurlock and Grover [2], effects of combustion-induced thermal expansion on turbulent flow were studied in a number of papers and various important phenomena were revealed, as reviewed elsewhere [3-6], see also recent Refs. [7-11]. In particular, a significant influence of thermal expansion on turbulent flow of unburned mixture upstream of a premixed flame was documented by analyzing Direct Numerical Simulation (DNS) data, with the effect being attributed to potential flow fluctuations caused by combustion-induced pressure perturbations upstream of the flame [12,13]. Moreover, analysis of the same DNS data has shown that flame-generated vorticity can mitigate an increase in the reaction-zone-surface area [14,15], contrary to the classical concept of combustion acceleration due to flame-generated turbulence. These recent results indicate that potential (irrotational) and solenoidal (rotational) velocity fluctuations differently affect the reaction zone and, thus, call for target-directed research into (a) characteristics of potential and solenoidal velocity fields, (b) their interplay within a premixed flame brush, and (c) influence of these two fields on the flame. However, these issues have not yet been investigated by the combustion community, to the best of the authors' knowledge, and the present work addresses this gap.

In particular, since the primary physical mechanism of the influence of turbulence on premixed combustion consists in stretching flames [16-18], contributions of solenoidal and potential velocity fields to the stretch rate

$$\dot{s} = a_t + 2S_d \kappa, \qquad (1)$$

which controls the rate of change of the local flame-surface area [17,18], are explored. Here, $a_t = -\mathbf{nn}:\mathbf{u} + \nabla \cdot \mathbf{u}$ is the strain rate, $\mathbf{u} = \{u_1, u_2, u_3\} = \{u, v, w\}$ is velocity vector, $\mathbf{n} = -\nabla c/|\nabla c|$ is the unit vector normal to an iso-surface of $c(\mathbf{x}, t) = $ const, $\kappa = \nabla \cdot \mathbf{n}/2$ is the iso-surface curvature, $S_d = [\nabla \cdot (\rho D \nabla c) + W]/(\rho |\nabla c|)$ is its displacement speed, $\rho$ is the density, $D$ is the molecular diffusivity of a combustion progress variable $c$, and $W$ is the mass rate of product creation. Since the term $S_d \kappa$ does not directly involve the velocity field (while the curvature is affected by the field), differences in contributions from solenoidal and potential velocity fields to the stretch rate are explicitly associated with differences in the solenoidal and potential strain rates. Accordingly, the latter differences are in the focus of the present study.

DNS data analyzed in the following are summarized in Section 2. Methods applied to process these data are reported in Section 3. Results are discussed in Section 4, followed by conclusions.

## 2. DNS Attributes

In the present study, DNS data computed almost 20 years ago [19,20] and processed in the aforementioned papers [12-15] are analyzed. The choice of this database, which appears to be outdated when compared to recent DNS data [9,21-24] generated invoking complex combustion chemistry at a high ratio of the rms turbulent velocity $u'$ to the laminar flame speed $S_L$, requires comments. The point is that the focus of the present study is placed on the influence of density variations on turbulent flows. Accordingly, combustion chemistry appears to be of secondary importance when compared to two other major requirements. First, to clearly reveal thermal expansion effects, data obtained at different density ratios $\sigma = \rho_u/\rho_b$ are required. The DNS data analyzed here were obtained at $\sigma = 2.5$ or 7.53. Second, a new tool (e.g. Helmholtz-Hodge decomposition of turbulent velocity field in a premixed flame) should initially be probed under conditions associated with the strongest manifestation of effects (e.g. thermal expansion effects) the tool aims at. Therefore, for the present goals, the flamelet regime of premixed turbulent burning [25] is of the most interest. The DNS by Nishiki et al. [19,20] did deal with the flamelet regime, as discussed in detail elsewhere [26], whereas the majority of recent DNS studies explored other combustion regimes.

Because the DNS data analyzed here were discussed in a number of papers by various research groups, e.g. see Refs. [12-15,19,20,26-28] and papers cited therein, we will restrict ourselves to a brief summary of those compressible simulations. They dealt with statistically 1D, planar, adiabatic flames modeled by unsteady 3D continuity, Navier-Stokes, and energy equations, supplemented with a transport equation for the mass fraction $Y$ of a deficient reactant and the ideal gas state equation. The Lewis and Prandtl numbers were equal to 1.0 and 0.7, respectively, and combustion chemistry was reduced to a single reaction. Accordingly, the combustion progress variable $c(\mathbf{x}, t) = 1 - Y(\mathbf{x},t)/Y_u = (T(\mathbf{x},t) - T_u)/(T_b - T_u)$. Here, subscripts $u$ and $b$ designate unburned and burned mixture, respectively. Temperature-dependence of molecular transport coefficients was taken into account, e.g. the kinematic viscosity $\nu = \nu_u (T/T_u)^{0.7}$.

The computational domain was a rectangular box $\Lambda_x \times \Lambda_y \times \Lambda_z$ with $\Lambda_x = 2\Lambda_y = 2\Lambda_z = 8$ mm, and was resolved using a uniform rectangular ($2\Delta x = \Delta y = \Delta z$) mesh of $512 \times 128 \times 128$ points. The flow was periodic in $y$ and $z$ directions. Homogeneous isotropic turbulence ($u' = 0.53$ m/s, an integral length scale $L = 3.5$ mm, and the turbulent Reynolds number $Re_t = u'L/\nu_u = 96$ [19,20]) was generated in a separate box and was injected into the computational domain through the left boundary. In the domain, the turbulence decayed along the direction $x$ of the mean flow.

At $t = 0$, a planar laminar flame was embedded into statistically the same turbulence assigned for the velocity field in the computational domain. Subsequently, the inflow velocity was increased twice, i.e., $U(0 \leq t < t_I) = S_L < U(t_I \leq t < t_{II}) < U(t_{II} \leq t) = S_t$, in order to keep the flame in the domain till the end $t_{III}$ of the simulations. Here, $S_t$ is the turbulent flame speed. Solely data obtained at $t_{II} \leq t \leq t_{III}$ are discussed here.

Two cases H and L characterized by High and Low, respectively, density ratios are investigated. In case H, $\sigma = 7.53$, $S_L = 0.6$ m/s, $\delta_L = 0.217$ mm, $Da = 18$, $Ka = 0.21$, $S_t = 1.15$ m/s. In case L, $\sigma = 2.5$, $S_L = 0.416$ m/s, $\delta_L = 0.158$ mm, $Da = 17.3$, $Ka = 0.30$, $S_t = 0.79$ m/s. Here, $\delta_L = (T_2 - T_1)/\max\{|\nabla T|\}$ is the laminar flame thickness, $Da = (L/u')/(\delta_L/S_L)$ and $Ka = (u'/S_L)^{3/2}(L/\delta_L)^{-1/2}$ are the Damköhler and Karlovitz numbers, respectively, evaluated at the leading edges of the mean flame brushes. The two flames are well associated with the flamelet combustion regime [25], e.g., various Bray-Moss-Libby (BML) expressions [29] hold, see figures 1-4 in Ref. [26].

When processing the DNS data, several types of mean quantities were calculated. First, stationary mean quantities $\bar{q}(x)$ were obtained by averaging the field $q(\mathbf{x}, t)$ over transverse planes $x =$const and time $t_{II} \leq t \leq t_{III}$. Stationary quantities $\bar{q}_u(x), \bar{q}_f(x), \bar{q}_r(x)$, and $\bar{q}_b(x)$ conditioned to unburned mixture, flame, reaction zone, and burned products, respectively, were sampled from points characterized by $c(\mathbf{x}, t) < 0.01, 0.01 \leq c(\mathbf{x}, t) \leq 0.99$, $0.75 < c(\mathbf{x}, t) < 0.97$ or $W[c(\mathbf{x}, t)] >$

max{$W(c)$}/2, and $0.99 < c(\mathbf{x},t)$, respectively, followed by averaging over transverse planes $x =$const and time $t_{II} \leq t \leq t_{III}$. Unsteady bulk quantities $\langle q \rangle_r(t)$ were sampled from the reaction zone ($0.75 < c(\mathbf{x},t) < 0.97$) over the entire computational domain. Finally, mean correlation coefficients for fields $q(\mathbf{x},t)$ and $p(\mathbf{x},t)$ were evaluated as follows

$$\bar{C}_{qp}(x) = \frac{\overline{qp} - \bar{q} \cdot \bar{p}}{\left(\overline{q^2} - \bar{q}^2\right)^{1/2} \left(\overline{p^2} - \bar{p}^2\right)^{1/2}} \tag{2}$$

and similar equations were applied to stationary conditioned coefficients $\left(\bar{C}_{qp}\right)_r(x)$ and unsteady bulk conditioned coefficients $\langle C_{qp} \rangle_r(t)$.

## 3. Decomposition of velocity field

If the velocity field $\mathbf{u}(\mathbf{x},t)$ is decomposed into solenoidal and potential subfields $\mathbf{u}_s(\mathbf{x},t)$ and $\mathbf{u}_p(\mathbf{x},t)$, respectively, then,

$$\nabla \times \mathbf{u}_s = \nabla \times \mathbf{u}, \qquad \mathbf{u}_p = \nabla \varphi, \qquad \mathbf{u} = \mathbf{u}_s + \mathbf{u}_p. \tag{3}$$

Such a decomposition is not unique unless the function $\varphi(\mathbf{x},t)$ is defined. To solve the problem, an extra constraint should be invoked and there are different methods for doing so. In the present work, two such methods, i.e. (i) widely-used "orthogonal" [30,31] and (ii) recently introduced "natural" [32,33] Helmholtz-Hodge decompositions, are adopted.

### 3.1. Orthogonal decomposition

The orthogonal decomposition invokes the following bulk constraint of orthogonality

$$\iiint_V \mathbf{u}_s \cdot \mathbf{u}_p d\mathbf{x} = 0 \tag{4}$$

of the subfields $\mathbf{u}_s(\mathbf{x},t)$ and $\mathbf{u}_p(\mathbf{x},t)$. Here, integration is performed over the computational domain $V$. The constraint results in the additivity of the bulk kinetic energies of the solenoidal and potential flow fields, i.e.

$$\iiint_V \mathbf{u} \cdot \mathbf{u} d\mathbf{x} = \iiint_V \mathbf{u}_s \cdot \mathbf{u}_s d\mathbf{x} + \iiint_V \mathbf{u}_p \cdot \mathbf{u}_p d\mathbf{x}. \tag{5}$$

Substitution of Eq. (3) into Eq. (4) yields

$$\iiint_V \mathbf{u}_s \cdot \nabla \varphi d\mathbf{x} = \iiint_V \nabla(\varphi \mathbf{u}_s) d\mathbf{x} - \iiint_V \varphi \nabla \cdot \mathbf{u}_s d\mathbf{x}$$

$$= \oiint_S \varphi \mathbf{u}_s \cdot \mathbf{n} dS - \iiint_V \varphi \nabla \cdot \mathbf{u}_s d\mathbf{x}, \tag{6}$$

where $S$ is the boundary of the domain $V$ and the unit vector $\mathbf{n}$ is normal to the boundary. If $\mathbf{u}_s \cdot \mathbf{n} = 0$ and $\nabla \cdot \mathbf{u}_s = 0$, then integrals in Eq. (6) vanish,

$$\Delta\varphi = \nabla \cdot \mathbf{u}_p = \nabla \cdot \mathbf{u} \qquad (7)$$

in the domain $V$, and

$$\left.\frac{\partial \varphi}{\partial n}\right|_S = \mathbf{n} \cdot \nabla\varphi|_S = \mathbf{n} \cdot \mathbf{u}|_S \qquad (8)$$

on its boundary. The Neumann problem given by Eqs. (7) and (8) has a unique solution for $\varphi(\mathbf{x}, t)$.

*3.2. Natural decomposition*

The natural decomposition [32,33] introduces an extra vector-field $\mathbf{w}(\mathbf{x}, t)$, which (i) coincides with the velocity field $\mathbf{u}(\mathbf{x}, t)$ in the domain $V$, i.e., $\mathbf{w}(\mathbf{x}, t) = \mathbf{u}(\mathbf{x}, t)$ for $\mathbf{x} \in V$, but (ii) is extrapolated to the entire 3D space $\mathbb{R}^3$ such that $|\mathbf{w}(\mathbf{x}, t)| \to 0$ for $|\mathbf{x}| \to \infty$. Subsequently, the velocity field $\mathbf{w}(\mathbf{x}, t)$ is decomposed as follows

$$\mathbf{w} = \nabla\Gamma + \nabla \times \mathbf{S}, \qquad \mathbf{x} \in \mathbb{R}^3. \qquad (9)$$

Consequently,

$$\Delta\Gamma = \nabla \cdot \mathbf{w}, \qquad \mathbf{x} \in \mathbb{R}^3, \qquad (10)$$

$$\nabla \times \nabla \times \mathbf{S} = \nabla \times \mathbf{w}, \qquad \mathbf{x} \in \mathbb{R}^3. \qquad (11)$$

Solutions to Poisson Eqs. (10) and (11) are unique

$$\Gamma(\mathbf{x}_0, t) = \iiint_{\mathbb{R}^3} G_\infty(\mathbf{x}, \mathbf{x}_0) \nabla \cdot \mathbf{w}(\mathbf{x}, t) d\mathbf{x}, \qquad \mathbf{x}_0, \mathbf{x} \in \mathbb{R}^3, \qquad (12)$$

$$\mathbf{S}(\mathbf{x}_0, t) = -\iiint_{\mathbb{R}^3} G_\infty(\mathbf{x}, \mathbf{x}_0) \nabla \times \mathbf{w}(\mathbf{x}, t) d\mathbf{x}, \qquad \mathbf{x}_0, \mathbf{x} \in \mathbb{R}^3. \qquad (13)$$

where $G_\infty(\mathbf{x}, \mathbf{x}_0) = -1/(4\pi|\mathbf{x} - \mathbf{x}_0|)$ is the free-space Green's function in $\mathbb{R}^3$. Finally, integration in Eqs. (12) and (13) is truncated outside the domain $V$ by interpreting the truncated integrals to be an external influence [33]. Thus,

$$\Gamma^*(\mathbf{x}_0, t) = \iiint_V G_\infty(\mathbf{x}, \mathbf{x}_0) \nabla \cdot \mathbf{u}(\mathbf{x}, t) d\mathbf{x}, \qquad \mathbf{x}_0, \mathbf{x} \in V, \qquad (14)$$

$$\mathbf{S}^*(\mathbf{x}_0, t) = -\iiint_V G_\infty(\mathbf{x}, \mathbf{x}_0) \nabla \times \mathbf{u}(\mathbf{x}, t) d\mathbf{x}, \qquad \mathbf{x}_0, \mathbf{x} \in V. \qquad (15)$$

Obviously, Eq. (3) with $\mathbf{u}_s = \nabla \times \mathbf{S}^*$ and $\mathbf{u}_p = \nabla\Gamma^*$ holds by virtue of Eqs. (9)-(11).

## 4. Results and Discussion

Figure 1 shows that, upstream of the mean flame brush (0.5 mm $< x <$ 1 mm), the solenoidal (curves 2 and 4) rms velocities $\overline{u_s'^2}$, $\overline{v_s'^2}$, and $\overline{w_s'^2}$, see the left, middle, and right columns, respectively, (i) are significantly larger than the potential (curves 1 and 3) rms velocities $\overline{u'^2_p}$, $\overline{v'^2_p}$, and $\overline{w'^2_p}$, respectively, and (ii) decay in the $x$-direction, as expected for constant-density turbulence. Within the mean H-flame brush (1 mm < $x$ < 4 mm), the decrease is reduced for $\overline{v'^2_s}$ and $\overline{w'^2_s}$, whereas $\overline{u'^2_s}$ significantly increases, see the top row in Fig. 1. This increase in $\overline{u'^2_s}$ is associated with vorticity generation by baroclinic torque, documented for flame H [27]. The potential rms velocities begin increasing already in the unburned mixture upstream of the mean flame brush ($x \approx 0.5$ mm), with $\overline{v'^2_p}$ and $\overline{w'^2_p}$ significantly exceeding $\overline{v'^2_s}$ and $\overline{w'^2_s}$, respectively, within the mean H-flame brush. In case L characterized by a low density ratio, such thermal-expansion effects are weakly pronounced, see the bottom row in Fig. 1, while the potential rms velocities still peak within the mean flame brush so that $\overline{u'^2_s}$ and $\overline{u'^2_p}$ or $\overline{w'^2_s}$ and $\overline{w'^2_p}$ are comparable there.

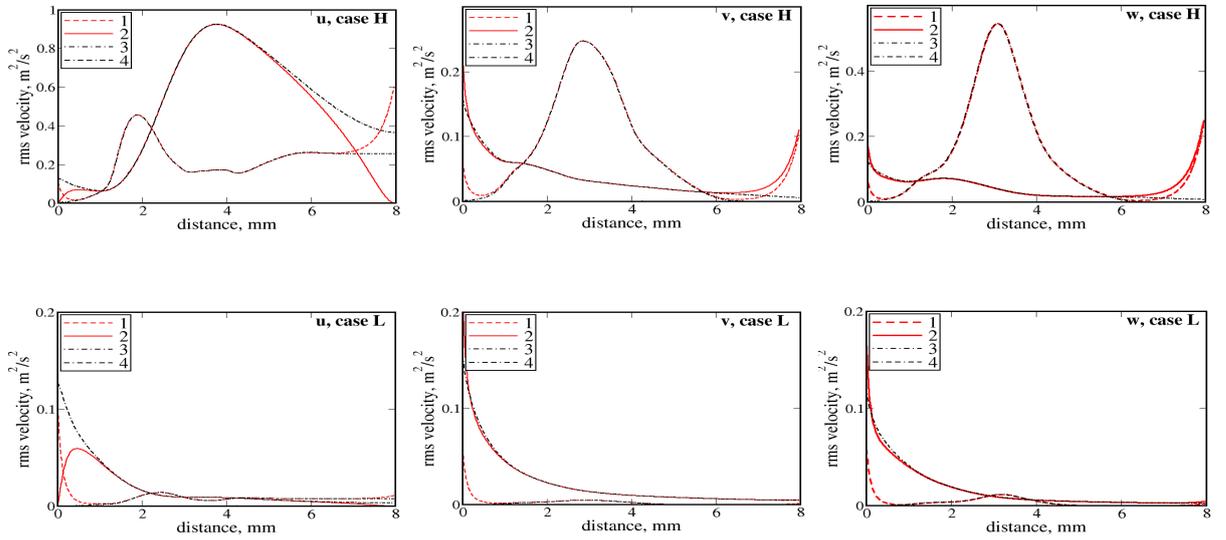

**Fig. 1.** Potential (curves 1 and 3) and solenoidal (curves 2 and 4) rms velocities $\overline{u'^2}$ (left column), $\overline{v'^2}$ (middle column), and $\overline{w'^2}$ (right column) obtained in case H (top row) or L (bottom row). Results obtained using orthogonal (natural) decomposition are shown in curves 1 and 2 (3 and 4, respectively).

Since comparison of curves 1 and 2 in Fig. 1 with curves 3 and 4, respectively, indicates that the two adopted Helmholtz-Hodge decompositions yield close results in the largest part of the computational domain except for regions near the inlet and outlet boundaries, we will restrict ourselves to presenting data obtained using the orthogonal decomposition. Figures created adopting the natural decomposition are hardly distinguishable from figures discussed in the following.

Coupling between the two velocity components is addressed in Figs 2 and 3. Figure 2 shows that while unconditioned correlations coefficients $\overline{\mathbf{u}'_p \cdot \mathbf{u}'_s} / [\overline{(\mathbf{u}'_p \cdot \mathbf{u}'_p)} \; \overline{(\mathbf{u}'_s \cdot \mathbf{u}'_s)}]^{1/2}$ are low within the mean flame brushes, the solenoidal and potential velocity fields conditioned to unburned gas (dashed lines) are well anti-correlated in the largest part of the H-flame brush and at the trailing edge of the L-flame brush. On the contrary, the two fields conditioned to burned gas (dotted-dashed lines) are well correlated in the largest part of the H-flame brush, but the effect is not pronounced in case L. The correlation coefficient conditioned to flame decreases from $\bar{c} = 0.2$ to $\bar{c} = 1$, see dotted lines, with the effect being more pronounced in case H.

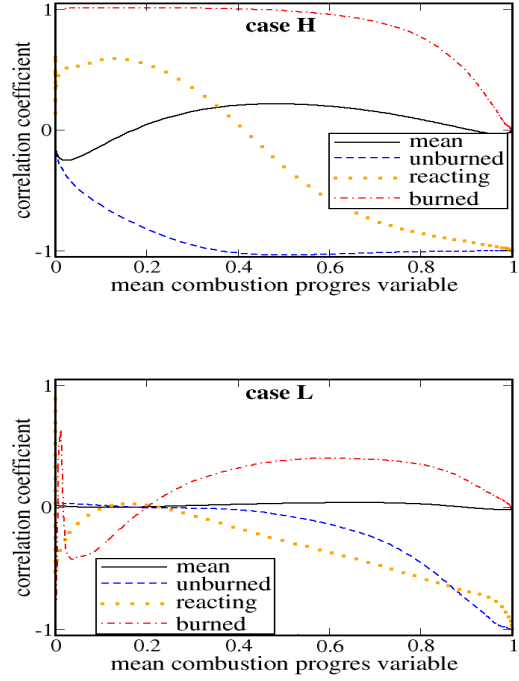

**Fig. 2.** Unconditioned (solid lines) and conditioned (broken lines) correlation coefficients $\overline{\mathbf{u}'_p \cdot \mathbf{u}'_s} / \left[ \overline{(\mathbf{u}'_p \cdot \mathbf{u}'_p)} \; \overline{(\mathbf{u}'_s \cdot \mathbf{u}'_s)} \right]^{1/2}$ vs. Reynolds-averaged combustion progress variable, computed in cases H and L.

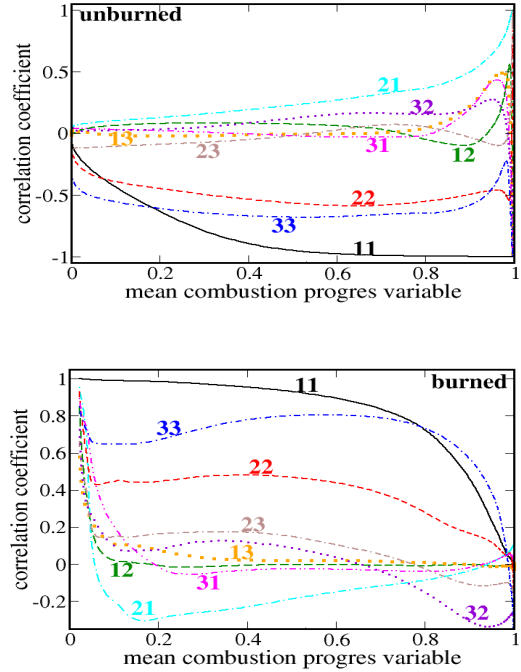

**Fig. 3.** Various components of the correlation tensors $\overline{u'_{i,p} u'_{j,s}} / \left( \overline{u'^2_{i,p}} \cdot \overline{u'^2_{j,s}} \right)^{1/2}$ conditioned to unburned or burned gas. Case H.

Figure 3 shows that the highly negative (positive) correlations $\overline{\mathbf{u}'_p \cdot \mathbf{u}'_s}$ conditioned to the unburned (burned, respectively) gas in the trailing (leading, respectively) halves of the H-flame brush are mainly controlled by correlations $\overline{u'_{1,p} u'_{1,s}}$ of the potential and solenoidal axial velocities. Negative (positive) correlations $\overline{u'_{2,p} u'_{2,s}}$ and $\overline{u'_{3,p} u'_{3,s}}$ of the transverse velocities conditioned to the unburned (burned, respectively) gas are also substantial in the trailing (leading, respectively) halves of the H-flame brush. Other components of the conditioned correlation tensors $\overline{u'_{i,p} u'_{j,s}}$ are small in the largest part of the H-flame brush. In case L (not shown), such effects are less pronounced, as implied in Fig. 2.

Let us consider strain rates $a_{t,p}$ and $a_{t,s}$ generated by the potential and solenoidal velocity fields, respectively. The aforementioned decrease in $\overline{(\mathbf{u}'_p \cdot \mathbf{u}'_s)}_f / \left[ \overline{(\mathbf{u}'_p \cdot \mathbf{u}'_p)}_f \overline{(\mathbf{u}'_s \cdot \mathbf{u}'_s)}_f \right]^{1/2}$ conditioned to flame with $\bar{c}$, see dotted lines in Fig. 2, manifests itself in a decrease in the correlation coefficient for $a_{t,p}$ and $a_{t,s}$, e.g. see Fig. 4.

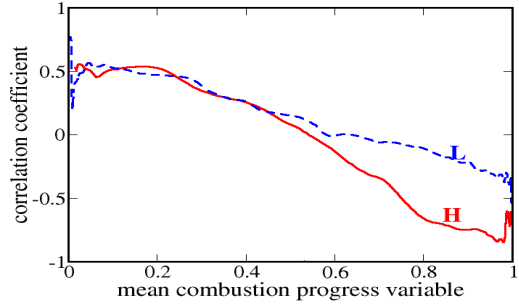

**Fig. 4.** Correlation coefficients for strain rates yielded by the potential and solenoidal velocity fields and conditioned to the reaction zones in case H (solid line) and L (dashed line).

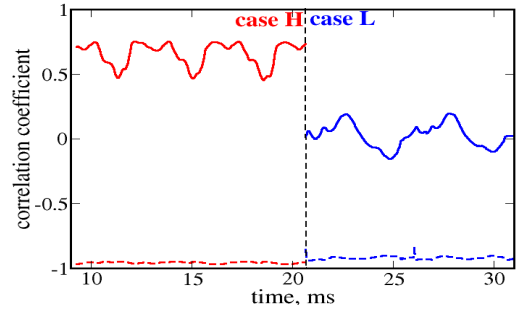

**Fig. 5.** Correlation coefficients between the local strain rate and curvature, conditioned to reaction zone and averaged over the entire volume of the flame brush vs. time. Solid and dashed lines show contributions from the solenoidal and potential velocity fields, respectively. Red and blue lines show results obtained in cases H and L, respectively.

Figures 5 and 6 show that the solenoidal and potential strain rates differently correlate with the local curvature. In both cases H and L, the bulk correlation coefficient $\langle C_{a_{t,p}\kappa} \rangle_r \approx -1$ for the potential flow in the entire flame brush at various instants, see dashed lines in Fig. 5. The stationary correlation coefficient $\left( \bar{C}_{a_{t,p}\kappa} \right)_r < -0.8$ in the entire H-flame brush and in the largest part of the L-flame brush, see dashed and dotted-dashed lines, respectively, in Fig. 6. $\langle C_{a_{t,s}\kappa} \rangle_r$ evaluated for the solenoidal flow oscillates around zero in case L, see blue solid line in Fig. 5, but this coefficient is positive in case H, see red solid line.

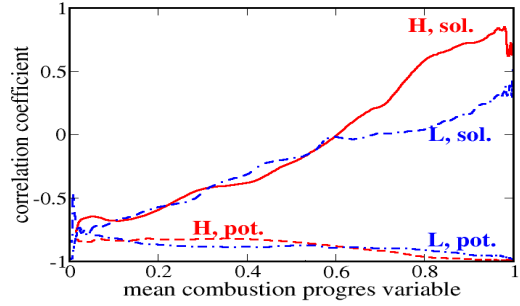

**Fig. 6.** Stationary correlation coefficients $\left(\bar{C}_{a_t\kappa}\right)_r$ between the local curvature and strain rate conditioned to the reaction zone vs. Reynolds-averaged combustion progress variable $\bar{c}$. Solid or double-dashed-dotted and dashed or dotted-dashed lines show contributions from the solenoidal and potential velocity fields, respectively. Red and blue lines show results obtained in cases H and L, respectively.

Negative correlation between flame strain rate and curvature was documented in various DNS studies of single-step and complex-chemistry premixed turbulent flames, e.g. see Refs. [34-36] and [37,38], respectively. The effect is commonly attributed to dilatation [36], whereas $a_t$ and $\nabla \cdot \mathbf{n}$ correlate weakly in constant-density turbulent reacting flows [39]. The present analysis offers an opportunity to consider the effect from another perspective. First, Figs. 5 and 6 imply that the negative correlation between $a_t$ and $\nabla \cdot \mathbf{n}$ results from strong negative correlation between the curvature and the strain rate $a_{t,p}$ created by the potential flow, which, in its turn, is strongly affected by combustion-induced thermal expansion in case H, see Fig. 1. Second, results obtained for the solenoidal velocity field in case L and plotted in Fig. 5 are consistent with the lack of a bulk correlation between $a_t$ and $\nabla \cdot \mathbf{n}$ in constant-density turbulent reacting flows [39]. Nevertheless, third, the solenoidal $a_{t,s}$ and $\nabla \cdot \mathbf{n}$ correlate negatively (positively) in the leading (trailing, respectively) halves of the L or H flame brush, see double-dashed-dotted and solid lines, respectively, in Fig. 6. In case L, the two halves of the flame brush well counterbalance one another, thus, yielding a low bulk correlation coefficient $\langle C_{a_{t,s}\kappa} \rangle_r$. In case H, such a mutual cancellation does not occur and the correlation coefficient $\langle C_{a_{t,s}\kappa} \rangle_r$ oscillates between 0.5 and 0.75, see red solid line in Fig. 5. Comparison of the positive $\langle C_{a_{t,s}\kappa} \rangle_r$ computed in case H with a small $\langle C_{a_{t,s}\kappa} \rangle_r$ calculated in case L implies that combustion-induced thermal expansion can significantly affect the correlation between the local curvature and the solenoidal strain rate, with the influence of the thermal expansion on the solenoidal $\langle C_{a_{t,s}\kappa} \rangle_r$ being opposite to the influence of the thermal expansion on the potential $\langle C_{a_{t,p}\kappa} \rangle_r$ in case H.

Figure 7 indicates that the solenoidal and potential strain rates not only correlate differently with the local reaction-zone curvature, but also have opposite signs (after averaging) in the largest part ($\bar{c} > 0.4$) of the H-flame brush, see red lines. Accordingly, the bulk conditioned potential (solenoidal) strain rate is positive (negative, respectively), see red dashed (solid, respectively) line in Fig. 8. In case L, such effects are less pronounced. Comparison of solid lines in Fig. 7 or 8 implies that combustion-induced thermal expansion affects not only potential, but also solenoidal strain rates. Indeed, both $\left(\overline{a_{t,s}|\nabla c|}\right)_r$ and $\langle a_{t,s}\rangle_r$ are close to zero in case L, but are negative in case H due to more pronounced thermal-expansion effects.

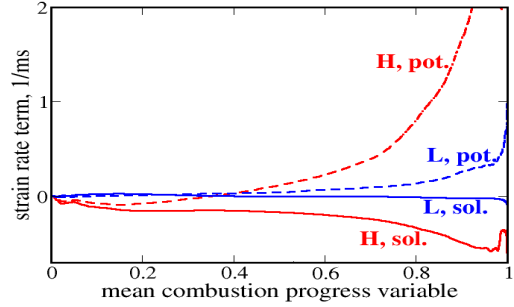

**Fig. 7.** Strain rate $\left(\overline{a_t|\nabla c|}\right)_r/\overline{|\nabla c|}_r$ conditioned to reaction zones. Solid and dashed lines show contributions from the solenoidal and potential velocity fields, respectively. Red and blue lines show results computed in cases H and L, respectively.

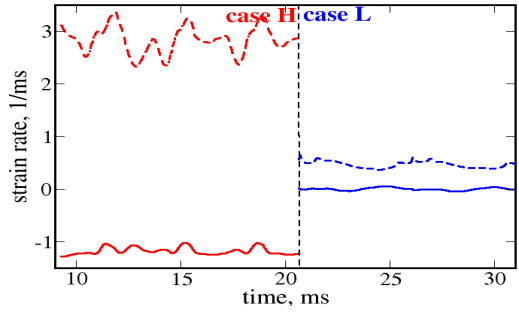

**Fig. 8.** Strain rates $\langle a_t\rangle_r$ conditioned to the reaction-zone and averaged over the entire volume of the flame brush vs. time. Legends are explained in caption to Fig. 7.

Finally, Fig. 9 shows that, in case H, the total stretch rate $\dot{s} = a_t + S_d\nabla\cdot\mathbf{n}$ is predominantly positive in regions characterized by a positive or a slightly negative ($a_{t,s}D_u/S_L^2 > -0.1$) solenoidal strain rate, see filled circles, whereas $\dot{s}$ is predominantly negative if the potential strain rate is positive and sufficiently large ($a_{t,p}D_u/S_L^2 > 0.3$), see open circles. Thus, the reaction-zone-surface area is decreased (increased) in regions characterized by large positive (negative) potential strain rates. Bearing in mind the well-pronounced negative correlation between $a_{t,p}$ and $\nabla\cdot\mathbf{n}$, see Figs. 5 and 6, the emphasized trend implies that the local stretch rate is controlled by the curvature term $S_d\nabla\cdot\mathbf{n}$ in such zones. In case L, the reaction-zone-surface area is increased (i.e. the probability of $\dot{s} > 0$ is significant) in regions characterized by low $a_{t,s}$ and $a_{t,p}$. Comparison of results computed in cases H ($\sigma = 7.53$) and L ($\sigma = 2.5$) shows that the density ratio substantially affects relationship between stretch rate within the reaction zone and potential or solenoidal local strain rate. Therefore, potential and solenoidal velocity fields affect the reaction-zone-surface area significantly different, with the effects being also qualitatively changed with variations in the density ratio.

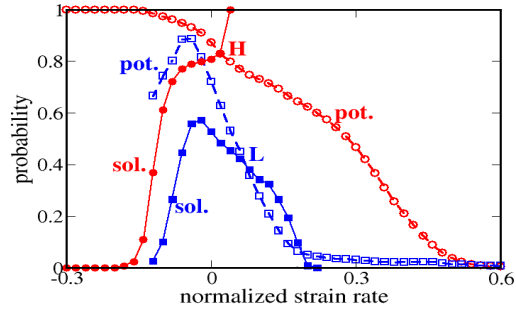

**Fig. 9.** Probabilities of positive local stretch rates conditioned to the reaction zone vs. the local strain rate due to solenoidal (solid lines with filled symbols) and potential (dashed lines with open symbols) velocity fields. The strain rate is normalized using $D_u/S_L^2$. Red and blue symbols show results obtained in cases H and L, respectively.

## 5. Conclusions

Two Helmholtz-Hodge decompositions were applied to processing DNS data obtained from two weakly turbulent premixed flames characterized by significantly different density ratios $\sigma$. Both decompositions yielded almost the same results within the mean flame brushes, thus, justifying such a research method. The following trends were documented. First, combustion-induced thermal expansion can significantly change turbulent flow of unburned mixture upstream of a premixed flame by generating potential velocity fluctuations, similarly to potential velocity perturbations generated by an unstable premixed flame in a laminar flow [40]. Second, the potential and solenoidal velocity fields are negatively (positively) correlated in unburned reactants (burned products, respectively) provided that the density ratio is large. Third, correlation between strain rates generated by the solenoidal and potential velocity fields and conditioned to the reaction zone is positive (negative) in the leading (trailing, respectively) halves of the mean flame brushes. Fourth, the potential strain rate correlates negatively with the curvature of the reaction zone, whereas the solenoidal strain rate and the curvature are negatively (positively) correlated in the leading (trailing, respectively) halves of the mean flame brushes. Five, in case H characterized by $\sigma = 7.53$, (i) the potential reaction-zone-conditioned strain rate is positive in the largest part of the mean flame brush whereas the solenoidal conditioned rate is negative; (ii) the bulk conditioned strain rate is positive (negative) for the potential (solenoidal) flow; and (iii) the stretch rate conditioned to the reaction zone is negative (positive) in regions characterized by large positive (negative) potential strain rates, whereas the opposite trend is observed for the solenoidal strain rate. Therefore, in case H, the potential and solenoidal velocity fields differently affect the reaction-zone-surface area. In case L characterized by $\sigma = 2.5$, such differences are significantly less pronounced.

**Acknowledgement**

VS gratefully acknowledges the financial support provided by ONERA and by the Grant of the Ministry of Education and Science of the Russian Federation (Contract No. 14.G39.31.0001 of 13.02.2017). AL gratefully acknowledges the financial support provided by CERC and Chalmers Area of Advance Transport.